\documentclass[10pt,twocolumn,twoside]{IEEEtran}
\IEEEoverridecommandlockouts

\usepackage{cite}
\usepackage{amsmath,amssymb,amsfonts}
\usepackage{graphicx}
\usepackage{textcomp}
\usepackage{balance}
\usepackage{xcolor}
\usepackage[utf8]{inputenc}
\usepackage{lipsum}
\usepackage{tikz}
\usetikzlibrary{positioning}
\usepackage{hyperref}
\usepackage{cleveref}
\usepackage{algorithm}
\usepackage[noend]{algpseudocode}
\usepackage{xspace}
\usepackage{mathtools}
\usepackage{booktabs} 
\usepackage{makecell}
\usepackage{multirow}
\usetikzlibrary{
    arrows.meta,                      % improved arrows
    patterns, patterns.meta,          % patterns support
    decorations, decorations.text, decorations.pathreplacing, 
    shapes, positioning, shadows, backgrounds, calc
}
\def\BibTeX{{\rm B\kern-.05em{\sc i\kern-.025em b}\kern-.08em
    T\kern-.1667em\lower.7ex\hbox{E}\kern-.125emX}}
\newcommand{\mb}[1]{\mathbf{#1}}
\newcommand{\mbg}[1]{\boldsymbol{#1}}

\newcommand{\cusum}{\text{CuSum}\xspace}

\title{\LARGE \bf 
Outage Identification from Electricity Market Data: \\Quickest Change Detection Approach
}

\author{%
  Milad Hoseinpour, 
  Shubhanshu Shekhar, and 
  Vladimir Dvorkin%
  \thanks{The authors are with the Department of Electrical Engineering and Computer Science, University of Michigan, Ann Arbor, MI 48109, USA. E-mail:\{miladh, shubhan, dvorkin\}@umich.edu.}% 
  \thanks{Supported by the Propelling Original Data Science (PODS) Program, Michigan Institute for Data and AI in Society (MIDAS). Project ID: 2521} 
}

\begin{document}
\begingroup
\allowdisplaybreaks

\maketitle

\begin{abstract}
Power system outages expose market participants to significant financial risk unless promptly detected and hedged. We develop an outage identification method from public market signals grounded in the parametric quickest change detection (QCD) theory. Parametric QCD operates on stochastic data streams, distinguishing pre- and post-change regimes using the ratio of their respective probability density functions. To derive the density functions for normal and post-outage market signals, we exploit multi-parametric programming to decompose complex market signals into parametric random variables with a known density. These densities are then used to construct a QCD-based statistic that triggers an alarm as soon as the statistic exceeds an appropriate threshold. Numerical experiments on a stylized PJM testbed demonstrate rapid line outage identification from public streams of electricity demand and price data.
\end{abstract}


\section{Introduction}

Power grids are facing increasing threats from extreme weather and aging infrastructure. Slow anomaly detection incurs enormous costs: power outages alone cost the U.S. economy as much as \$150 billion annually \cite{pew2015electricgrid}. The outages caused by the 2021 Winter Storm Uri led to billions in damages in Texas, with days-long blackouts, and surging prices up to \$9,000/MWh \cite{ferc2021february}. Early warning signs appeared in public price signals, yet delayed detection and hedging drove major utilities into bankruptcy \cite{li2024extreme}. While subsequent reforms mandated faster outage reporting by ERCOT \cite{ercot2021discussion}, market participants (utilities, traders, etc.) require independent detection  to act preemptively rather than await official reports.

System operators across the U.S. publish information on marginal prices, aggregated generator and load statistics, binding constraints, along with other operational and market data. This has given rise to modern data analytics companies, such as Grid Status \cite{gridstatus} and Electricity Maps \cite{electricitymaps_live}, which provide real-time access to market data streams, leading to the question: \textit{How to leverage public operational and market data for independent and fast anomaly detection in power systems?}

The theory of Quickest Change Detection (QCD) offers the right formalism to answer this question \cite{veeravalli2014quickest}. QCD involves continuously monitoring a data stream, e.g., electricity prices, to raise an alarm as soon as an abrupt distributional change is detected.  A Cumulative Sum (\cusum) is a powerful QCD statistic that accumulates evidence over time and compares it against a threshold, providing a theoretically optimal framework for detecting abrupt distributional changes~\cite{page1954continuous}. These properties have led to the development of numerous variants of \cusum informed by the network topology~\cite{chen2015quickest,rovatsos2017statistical,nizam2016attack,raghavan2010quickest}. However, topology alone may not fully characterize the cause of distributional shifts. This is particularly relevant to electricity markets governed by optimization engines with specific objectives and constraints. For example, a transmission line outage in power systems triggers an economic re-dispatch optimization, whose solution must respect generation cost structure, along with power flow and generation limits in the post-outage topology. This raises the question: \textit{How to integrate optimization objectives and constraints within the design of \cusum statistics for faster outage detection?}

\underline{Contributions}: We propose a \cusum-based method for detection and identification of critical events (e.g., line or generator outage) in power systems using publicly available market data streams. Our method is based on the statistical relationship between observable market data (e.g., demand, prices, and aggregated dispatch statistics) and the underlying market-clearing structure, that is leveraged in a \cusum scheme accumulating evidence of structural changes in real-time. We contribute to anomaly detection in power systems by: 
\begin{enumerate}
    \item Developing probabilistic models of real-time market data streams in normal and post-outage states using multi-parametric programming \cite{tondel2003algorithm,taheri2020fast,taheri2021strategic}. The models decompose stochastic signals into simpler streams that exhibit distinct characteristics in normal and post-outage states, facilitating outage detection and identification.
    \item Designing a \cusum-based statistic that leverages these models to identify structural changes, such as line and generator outages, from streams of market-clearing data. The proposed statistic tests whether the observed stream originates from the normal or post-outage model, accumulating statistical evidence when an outage occurs and remaining quiescent otherwise.
    \item Identifying and exploring detection thresholds for the proposed statistic, enabling the control of false alarm rates and detection delays. 
\end{enumerate}

We apply the method to line outage identification using electricity demand and prices from the stylized PJM testbed.

\underline{Related Work}: Outage detection is central to power system operations. A substantial body of work has focused on outage identification from streams of Phasor Measurement Unit (PMU) data, with early contributions relating observed streams to those induced by models under varying topologies \cite{tate2008line,tate2009double}. The combinatorial nature of line outage identification subsequently led to mixed-integer formulations, as in \cite{emami2012external}. The computational burden was later alleviated by framing the problem as sparse recovery problem in \cite{zhu2012sparse} and by leveraging related methods from the signal processing domain \cite{giannakis2013monitoring}. More recently, detection has been offloaded to machine learning models, e.g., based on convolutional neural networks \cite{li2019real} and decision trees \cite{hannon2021real}. Despite strong empirical validation, these methods lack awareness of the fundamental performance metrics in stochastic settings: detection delay and false alarm rate. The two are intrinsic to QCD schemes, which have been applied to line outage detection from PMU measurements in both quasi-steady-state \cite{chen2015quickest} and transient \cite{rovatsos2017statistical} settings. Existing QCD schemes, however, rely on PMU measurements available only to system operators, rendering them unsuitable for \textit{independent} analytics from streams of market-clearing data.

The use of public market data for inference was explored in \cite{kekatos2014grid,kekatos2015online} using electricity prices for topology identification. In a similar spirit, inverse optimization techniques have been applied to recover market structure, including transmission constraints \cite{birge2017inverse}, supply curves \cite{mitridati2017bayesian,liang2023data}, and electricity demands \cite{dvorkin2020differentially}. While able to extract structural information, these methods assume \textit{static} market structure. Hence, they do not provide definitive outage detection, as they fail to distinguish between pre- and post-outage samples in data streams. When applied to outage identification (e.g., inferring that a faulted line has zero transfer capacity), the inertia induced by pre-outage samples leads to significant detection delays.

By contrast, the proposed QCD statistic posits nominal and candidate post-outage market structures (e.g., one per line outage) and acts on public market signals to raise an alarm as soon as possible after a structural change occurs. We focus on parametric QCD \cite{xie2021sequential}, which requires explicit probabilistic densities of observed signals. This is not restrictive for market-clearing inputs such as electricity demand and renewable power generation, which are known to be well-approximated by parametric distributions \cite{roald2023power}. However, the densities of dispatch statistics and electricity prices are not readily available, as market-clearing optimization is a nonlinear transformation of input parameters. To address this challenge, we exploit multi-parametric programming \cite{tondel2003algorithm,taheri2020fast,taheri2021strategic}, which decomposes market-clearing optimization into critical regions, each admitting an affine mapping from inputs to market outcomes. The affine structure of critical regions permits closed-form derivation of output densities from known input distributions, thus enabling parametric QCD on public market signals.

\underline{Paper Organization}: Next Section~\ref{stat_model} details probabilistic models of market-clearing data streams, which are then leveraged in  Section~\ref{CUSUM_method} to design the \cusum detection statistics. Section~\ref{numerical_results} reports numerical results, and Section~\ref{conclusion} concludes. All auxiliary information is relegated to the Appendix~\ref{app:qp_params}. 

\section{Probabilistic Models of Real-Time Market}\label{stat_model}
In this section, we derive probabilistic models of observed data streams from system operators' dashboards, including locational marginal prices (LMPs) and aggregated dispatch statistics. We begin with a perturbed market-clearing problem that is repeatedly solved under stochastic demand realizations, and then leverage multi-parametric programming theory \cite{tondel2003algorithm} to obtain tractable models that allow us to characterize the probability density functions of observations.

\subsection{Perturbed Market-Clearing Optimization Problem}
We consider electricity market-clearing problem, which is solved every 5 minutes for real-time realizations of demand $\mb{l}+\mbg{\xi}$, modeled by the mean vector $\mb{l}$ perturbed by random vector $\mbg{\xi}$. In response to new demand, system operators compute the cost-optimal dispatch $\mb{p}$ of conventional generators subject to generation and power flow constraints. When conventional generation  is insufficient to meet demand, system operators resort to load shedding, denoted by $\mbg{\ell}$. The perturbed market-clearing is posed as the following convex problem\cite{chatzivasileiadis2018lecture}:
\begin{subequations}\label{eq:PDCOPF}
\begin{align}
\underset{\mb{p},\mbg{\ell}}{\text{minimize}}\quad 
& \tfrac{1}{2} \mb{p}^\top \mb{C} \mb{p} + \mb{c}^\top \mb{p}
  + \tfrac{1}{2} \mbg{\ell}^\top \mb{S} \mbg{\ell} + \mb{s}^\top \mbg{\ell}
\label{eq:PDCOPF_obj} \\
\text{subject to}\quad
& \mb{1}^{\top}\big(\mb{M}_{\text{p}} \mb{p} - \mb{M}_{\text{l}}(\mb{l} + \mbg{\xi} - \mbg{\ell})\big) = 0,
\label{eq:PDCOPF_balance} \\
& |\mb{F}\big(\mb{M}_{\text{p}} \mb{p} - \mb{M}_{\text{l}}(\mb{l} + \mbg{\xi} - \mbg{\ell})\big)| \leqslant \bar{\mb{f}},
\label{eq:PDCOPF_flow_pos} \\
& \underline{\mb{p}} \leqslant \mb{p} \leqslant \overline{\mb{p}},
\label{eq:PDCOPF_gen_up} \\
& \mb{0}\leqslant\mbg{\ell} \leqslant \mb{l} + \mbg{\xi},
\label{eq:PDCOPF_load_up}
\end{align}
\end{subequations}
where objective function~\eqref{eq:PDCOPF_obj} minimizes quadratic costs of generation, with parameters $\mb{C}$ and $\mb{c}$, and load shedding, parameterized by $\mb{S}$ and $\mb{s}$. Constraint~\eqref{eq:PDCOPF_balance} enforces power balance and \eqref{eq:PDCOPF_flow_pos} introduces power flow limits using a power transfer distribution factor (PTDF) matrix $\mb{F}$. The incidence matrices $\mb{M}_{\text{p}}$ and $\mb{M}_{\text{l}}$ relate generators and demands to their corresponding buses, respectively. Generator limits are enforced by~\eqref{eq:PDCOPF_gen_up}, while constraints~\eqref{eq:PDCOPF_load_up} bound load shedding between zero and real-time demand realizations.

For further derivation, it is convenient to rewrite the perturbed problem~\eqref{eq:PDCOPF} in a compact quadratic programming (QP) form. Let $\mb{x} := [\,\mb{p}^{\top},\,\boldsymbol{\ell}^{\top}\,]^{\top}$ denote the vector of decision variables. Then, the problem can be written as
\begin{subequations}\label{prob:qp}
\begin{align}
   \underset {\mb{x}}{\text{minimize}}\quad
    & \tfrac{1}{2}\mb{x}^{\top}\mb{Q}\mb{x} + \mb{q}^{\top}\mb{x} \label{prob:qp_obj}\\
    \text{subject to}\quad
    & \mb{A}\mb{x} \leqslant\mb{B}\boldsymbol{\xi} + \mb{b} 
    \quad : \mbg{\mu},\label{prob:qp_con}
\end{align}
\end{subequations}
where vector $\boldsymbol{\mu}$ stacks the duals, from which the LMPs $\mbg{\lambda} = \mbg{\Lambda}\mbg{\mu}$ are obtained using auxiliary matrix $\mbg{\Lambda}$. Parameters $\mb{Q}$, $\mb{q}$, $\mb{A}$, $\mb{B}$, $\mb{b},$ and $\mbg{\Lambda}$ are defined in Appendix~\ref{app:qp_params}.

\subsection{Mapping Load Perturbations to Market Outcomes}
The uncertainty of demand $\mbg{\xi}$ is often described by known probability distributions, e.g., Normal, Log-Normal, or Weibull \cite{roald2023power}. However, characterizing the distributions of aggregated dispatch $\mb{1}^{\top}\mb{x}$ and prices $\mbg{\lambda}$ via density of $\mbg{\xi}$ is challenging since these quantities depend nonlinearly on $\mbg{\xi}$ through optimization \eqref{prob:qp}. We overcome this challenge using multi-parametric programming \cite{tondel2003algorithm}, providing an adaptive affine mapping between load perturbations and market outcomes.

Consider some realization of the demand perturbation $\Hat{\mbg{\xi}}$, and  let $\tilde{\mb{A}}, \tilde{\mb{B}}, \tilde{\mb{b}}$ contain those entries of ${\mb{A}}, {\mb{B}}, {\mb{b}}$ that correspond to the active constraints at $\Hat{\mbg{\xi}}$, and let $\overline{\mb{A}}, \overline{\mb{B}}, \overline{\mb{b}}$ denote the remaining entries corresponding to inactive constraints. The stationarity and primal feasibility conditions of \eqref{prob:qp} yield 
\begin{subequations}
\begin{align}
\mb{Q}\mb{x}^\star
+ \mb{q}
+ \tilde{\mb{A}}^{\top}\tilde{\mbg{\mu}}^\star
&= \mb{0}, \label{eq:kkt_stationarity}\\
\tilde{\mb{A}}\mb{x}^\star
- \tilde{\mb{B}}\Hat{\mbg{\xi}}
- \tilde{\mb{b}}
&= \mb{0}, \label{eq:kkt_primal}
\end{align}
\end{subequations}
at optimality. Solving~\eqref{eq:kkt_stationarity} for $\mb{x}^{\star}$ and substituting into~\eqref{eq:kkt_primal}, we obtain an affine relationship between the dual solution and the demand perturbation:
\begin{subequations}
\begin{align}
\tilde{\mbg{\mu}}^\star(\Hat{\mbg{\xi}})
=
\mb{D}\Hat{\mbg{\xi}}
+
\mb{d},
\label{eq:mu_affine}
\end{align}
with auxiliary parameters $\mb{D}$ and $\mb{d}$ defined as
\begin{align}
\mb{D}
&:=
-
\left(
\tilde{\mb{A}}\mb{Q}^{-1}\tilde{\mb{A}}^{\top}
\right)^{-1}
\tilde{\mb{B}},
\label{eq:D_def}\\
\mb{d}
&:=
-
\left(
\tilde{\mb{A}}\mb{Q}^{-1}\tilde{\mb{A}}^{\top}
\right)^{-1}
\left(
\tilde{\mb{b}}
+
\tilde{\mb{A}}\mb{Q}^{-1}\mb{q}
\right).
\label{eq:d_def}
\end{align}
\end{subequations}
Substituting \eqref{eq:mu_affine} into  \eqref{eq:kkt_stationarity} and solving for $\mb{x}^{\star}$, the corresponding primal solution also becomes affine in the perturbation
\begin{subequations}\label{eq:x_affine}
\begin{align}
\mb{x}^\star(\Hat{\mbg{\xi}})
&=
\mb{P}\Hat{\mbg{\xi}}
+
\mb{p}, \label{eq:x_affine:a}
\end{align}
with auxiliary parameters $\mb{P}$ and $\mb{p}$ defined as
\begin{align}
\mb{P}
&:=
-\mb{Q}^{-1}\tilde{\mb{A}}^{\top}\mb{D}, \label{eq:P_def}\\
\mb{p}
&:=
-\mb{Q}^{-1}\mb{q}
-\mb{Q}^{-1}\tilde{\mb{A}}^{\top}\mb{d}. \label{eq:p_def}
\end{align}
\end{subequations}
Taken together, for some realization $\Hat{\mbg{\xi}}$  of the random perturbation, the primal-dual solution depends affinely on~$\Hat{\mbg{\xi}}$:
\begin{align}
\begin{bmatrix}
\mb{x}^\star\\
\tilde{\mbg{\mu}}^\star
\end{bmatrix}\!(\Hat{\mbg{\xi}})
=
\begin{bmatrix}
\mb{P}\\
\mb{D}
\end{bmatrix}
\Hat{\mbg{\xi}}
+
\begin{bmatrix}
\mb{p}\\
\mb{d}
\end{bmatrix}.
\label{eq:affine_solution}
\end{align}

\begin{algorithm}[b]
\caption{Computing critical regions of problem \eqref{prob:qp}}
\label{alg:critical_regions_nominal}
\begin{algorithmic}[1]
\Statex \textbf{Input:} Market-clearing parameters $\mb{Q}$, $\mb{q}$, $\mb{A}$, $\mb{B}$, and $\mb{b}$, and load perturbation set $\Xi$
\Statex \textbf{Output:} Set of critical regions $\mathcal{CR}$ 
\State Initialize $\mathcal{CR} \gets \emptyset$
\For{each load perturbation $\Hat{\mbg{\xi}} \in \Xi$}
    \If{$\Hat{\mbg{\xi}}$ lies in an existing region $\mathcal{R}_i\in\mathcal{CR}$}
        \State \textbf{skip} \Comment{Already covered by an existing region}
    \EndIf
    \State Solve the QP problem~\eqref{prob:qp} at $\mbg{\xi}=\Hat{\mbg{\xi}}$
    \State Identify active and inactive constraints 
    \State Compute the critical region $\mathcal{R}_i$ via \eqref{set:critical_region} and add to $\mathcal{CR}$
\EndFor

\State \textbf{return} $\mathcal{CR}$
\end{algorithmic}
\end{algorithm}

An important result from \cite{tondel2003algorithm} is that the affine mapping~\eqref{eq:affine_solution} is valid not only for a particular realization $\Hat{\mbg{\xi}}$, but for all $\mbg{\xi}$ from the polyhedral neighborhood of $\Hat{\mbg{\xi}}$, referred to as \textit{critical region}. The critical region $\mathcal{R}$ of $\Hat{\mbg{\xi}}$ is defined by the primal feasibility conditions on inactive constraints and the dual feasibility conditions on active constraints:
\begin{align}
    \mathcal{R}=\left\{\mbg{\xi}\;|\; (\overline{\mb{A}}
        \mb{P} - \overline{\mb{B}})\mbg{\xi}
        < 
        \overline{\mb{b}} - \overline{\mb{A}}\mb{p},\; \mb{D}\mbg{\xi} + \mb{d} \geqslant \mb{0}\right\}.\label{set:critical_region}
\end{align}

A well-posed QP \eqref{prob:qp} has a finite number of critical regions \cite{tondel2003algorithm}, each having its own affine mapping of load perturbations to optimization results. Hence, observing perturbation $\Hat{\mbg{\xi}}$, we pick the corresponding critical region and map the density of $\Hat{\mbg{\xi}}$ to the density of market outcomes via \eqref{eq:affine_solution} from that region.

The set $\mathcal{CR}$ of critical regions is computed using Alg. \ref{alg:critical_regions_nominal}. It takes market-clearing parameters and the set $\Xi$ of perturbation samples as inputs. For each sample, it solves the QP problem \eqref{prob:qp} and computes the associated region, unless the sample belongs to one of the regions found at previous iterations. 

\subsection{Probabilistic Models of Electricity Market Data Streams}

The multi-parametric decomposition of problem \eqref{prob:qp} allows us to derive densities of market-clearing outcomes reported in system operators' dashboards through adaptive transformations of the demand perturbation density. When perturbations belong to critical region $\mathcal{R}_i$, market outcomes follow region-specific affine relationship with this density \cite[Theorem 2.1.8]{casella2001statisticalinference}. For example, LMPs can be expressed using~\eqref{eq:affine_solution} as
\begin{align}
    \mbg{\lambda}_{i}(\mbg{\xi}) = \tilde{\mbg{\Lambda}}_{i}\tilde{\mbg{\mu}}_{i}^{\star}(\mbg{\xi}) = \tilde{\mbg{\Lambda}}_{i}\mb{D}_{i}\mbg{\xi} + \tilde{\mbg{\Lambda}}_{i}\mb{d}_{i},\label{eq:lmp_affine}
\end{align}
where $\mb{D}_{i}$ and $\mb{d}_{i}$ are parameters of the critical region $\mathcal{R}_i$, and matrix $\tilde{\mbg{\Lambda}}_{i}$ discards the columns of the full matrix $\mbg{\Lambda}$ corresponding to inactive power flow constraints in region $i$. When a critical region changes over time, expression \eqref{eq:lmp_affine} adapts parameters $\mb{D}_{i},\mb{d}_{i},$ and $\tilde{\mbg{\Lambda}}_{i}$ to a new critical region.

Before deriving the exact density, we note that the classical \cusum method assumes i.i.d. observations \cite{basseville1995detection}, an assumption that rarely holds in market data streams due to demand seasonality (e.g., daily morning and evening peaks). To satisfy \cusum assumptions, we act on the increments of observed signals, thereby producing approximately stationary observations with tractable distributions. The increment of LMPs are
\begin{align}
    \Delta\mbg{\lambda}_{t} = \mbg{\lambda}_{i}(\mbg{\xi}_{t}) - \mbg{\lambda}_{i}(\mbg{\xi}_{t-1}) ,
\end{align}
computed at two subsequent time steps.

Suppose that the increments of demand perturbations follow a Wiener process, i.e., $\mbg{\xi}_t - \mbg{\xi}_{t-1} \sim \mathcal{N}(\mb{0},\mb{\Sigma})$, with zero-mean and known covariance matrix $\mb{\Sigma}$ estimated from operational history. Then, the mean and covariance of LMP increments $\Delta\mbg{\lambda}_{t}$ from the critical region $\mathcal{R}_i$ are given by
\begin{align}
\mathbb{E}[\Delta\mbg{\lambda}_{t}]
&=
\tilde{\mbg{\Lambda}}_{i}\mb{D}_{i} \, \mathbb{E}[\mbg{\xi}_t - \mbg{\xi}_{t-1}]
=
\mb{0}, \\
\mathrm{Cov}[\Delta\mbg{\lambda}_{t}]
&=
\mathrm{Cov}\!\left[\tilde{\mbg{\Lambda}}_{i}\mb{D}_{i} (\mbg{\xi}_t - \mbg{\xi}_{t-1})\right]
=
\tilde{\mbg{\Lambda}}_{i}\mb{D}_{i} \mb{\Sigma} (\tilde{\mbg{\Lambda}}_{i}\mb{D}_{i})^\top. \notag
\end{align}

Accordingly, within each critical region $\mathcal{R}_i$, the incremental observation $\Delta\mbg{\lambda}_t$ follows a zero-mean multivariate Gaussian distribution with probability density function proportional to
\begin{align}
f_i(\Delta\mbg{\lambda}_t)
\propto
\exp\!\left(
-\tfrac{1}{2}
\Delta\mbg{\lambda}_t^\top
(\tilde{\mbg{\Lambda}}_{i}\mb{D}_{i} \mb{\Sigma} (\tilde{\mbg{\Lambda}}_{i}\mb{D}_{i})^\top)^{-1}
\Delta\mbg{\lambda}_t
\right).\label{eq:general_pdf_price}
\end{align}

With a finite set $\mathcal{CR}=\{\mathcal{R}_{i}\}_{i=1}^{I}$ of critical regions, the probabilistic model of real-time LMPs is given by a finite collection of probability density functions \eqref{eq:general_pdf_price}.

In what follows, we will narrow down the scope of observed market signals to LMPs alone. This is without much loss of generality, as we can derive the densities for other published statistics (total generation by fuel type, emissions, power flows in tie-lines, etc.) in a similar way, as long as they admit an affine transformation of the primal-dual solution in \eqref{eq:affine_solution}. For example, the density for increments of the total generator dispatch $g_{i}(\mbg{\xi})=\mb{1}^{\top}\mb{x}_{i}^{\star}(\mbg{\xi})$ takes the following form 
\begin{align}
f_i(\Delta g_t)
\propto
\exp\!\left(
-\tfrac{1}{2}
\Delta g_t
(\mb{1}^{\top}\mb{P}_{i} \mb{\Sigma} (\mb{1}^{\top}\mb{P}_{i})^\top)^{-1}
\Delta g_t
\right),\label{eq:general_pdf_dispatch}
\end{align}
which is analogous to \eqref{eq:general_pdf_price}, except that it adapts the primal parameters $\mb{P}_{i}$ of critical region $\mathcal{R}_i$ to demand realizations.

\section{\cusum-Based Outage Identification from Streams of Market-Clearing Data}\label{CUSUM_method}

Consider a stream $\Delta\mbg{\lambda}_1,\Delta\mbg{\lambda}_2,\Delta\mbg{\lambda}_3,\dots$ of LMP increments as a stochastic process whose statistical properties are determined by the underlying market-clearing structure. Under normal operating conditions, the system evolves according to a nominal structure, and the distribution of $\Delta\mbg{\lambda}_t$ remains stationary over time. A structural change, such as a line or generator outage, alters the feasible set of the market-clearing problem, leading to an abrupt shift in LMP distribution. Formally, let $\mathcal{A}$ denote the set of possible structural changes in the market-clearing problem. The family of distributions $\{f^a\}_{a\in\mathcal{A}\cup\{0\}}$ is defined such that $a=0$ corresponds to the nominal structure, and $a\in\mathcal{A}$ corresponds to a particular structural change. We assume the existence of an unknown change (outage) point $T\in\mathbb{N}\cup\{\infty\}$ such that
\begin{equation}
\Delta\mbg{\lambda}_{t} \sim
\begin{cases}
f^0, & t \leqslant T, \\
f^a, & t > T,
\end{cases}
\qquad a\in\mathcal{A}.
\end{equation}

Our goal is twofold: (1) promptly detect the structural change from the streams of LMP observation, and (2) identify the specific outage underlying the change. That is, we need to distinguish between the null hypothesis $\mathbb{H}_0:T=\infty$ (no change), and a set of alternative $\mathbb{H}_1:T<\infty$ hypotheses using streams of market observations. Toward this goal, in this section we develop  \cusum statistics where the log-likelihood ratios are computed using probabilistic market models \eqref{eq:general_pdf_price} induced by time-varying critical regions.

\subsection{\cusum Design using Probabilistic Market Models}
Consider two variants of market-clearing problem~\eqref{prob:qp}, which are a nominal problem modeling normal operation: 
\begin{subequations}\label{prob:nominal}
\begin{align}
    \underset{\mb{x}}{\text{minimize}}\quad
    &\tfrac{1}{2}\mb{x}^{\top}\mb{Q}_{0}\mb{x} + \mb{q}_{0}^{\top}\mb{x} \\
    \text{subject to}\quad
    &\mb{A}_{0}\mb{x} \leqslant \mb{B}_{0}\mbg{\xi} + \mb{b}_{0}
    \quad:\mbg{\mu}_{0},
\end{align}
\end{subequations}
and an alternative problem corresponding to an outage:
\begin{subequations}\label{prob:alternative}
\begin{align}
    \underset{\mb{x}}{\text{minimize}}\quad
    &\tfrac{1}{2}\mb{x}^{\top}\mb{Q}_{a}\mb{x} + \mb{q}_{a}^{\top}\mb{x} \\
    \text{subject to}\quad
    &\mb{A}_{a}\mb{x} \leqslant \mb{B}_{a}\mbg{\xi} + \mb{b}_{a}
    \quad:\mbg{\mu}_{a}.
\end{align}
\end{subequations}

A line outage is modeled by excluding the entries associated with the
faulted line from $\mb{A}_{0}$, $\mb{B}_{0}$, and $\mb{b}_{0}$, yielding $\mb{A}_{a}$, $\mb{B}_{a}$, and $\mb{b}_{a}$. A generation outage is modeled by removing the relevant entry in $\mb{x}$ along with related problem data. The nominal and alternative problems have distinct critical regions and thus distinct LMP densities \eqref{eq:general_pdf_price}, facilitating outage detection.

We propose a two-stage \cusum design. At the first (offline) stage, before observing market signals, we compute the critical regions for the nominal and alternative market structures, denoted by $\mathcal{CR}_{0} = \{\mathcal{R}_{i}^{0}\}_{i=1}^{I}$ and $\mathcal{CR}_{a} = \{\mathcal{R}_{j}^{a}\}_{j=1}^{J}$, respectively, using Alg.~\ref{alg:critical_regions_nominal}. At the second (online) stage, at each time step $t$ we observe LMP increments $\Delta\mbg{\lambda}_{t}$ and the corresponding demand perturbation $\mbg{\xi}_{t}$. Based on realizations of $\mbg{\xi}_{t}$, we identify the critical region $\mathcal{R}_{i}^{0}\in\mathcal{CR}_{0}$ for the nominal problem and $\mathcal{R}_{j}^{a}\in\mathcal{CR}_{a}$ for all alternative problems. The log-likelihood ratio for testing whether increments come from the nominal or alternative problem is computed as: 
\begin{align}
&\log\!\left(\frac{f_j^{a}(\Delta\mbg{\lambda}_{t})}{f_i^{0}(\Delta\mbg{\lambda}_{t})}\right)=
\nonumber\\
&\;=
\log\!\left(
\!\frac{
\exp\!\left(
\!-\frac{1}{2}\Delta\mbg{\lambda}_{t}^{\top}\!
\Big(\tilde{\mbg{\Lambda}}_{j}^{a}\mb{D}_{j}^{a} \mb{\Sigma} (\tilde{\mbg{\Lambda}}_{j}^{a}\mb{D}_{j}^{a})^\top\!\Big)^{-1}\!\!
\Delta\mbg{\lambda}_{t}\!
\right)
}{
\exp\!\left(
\!-\frac{1}{2}\Delta\mbg{\lambda}_{t}^{\top}\!
\Big(\tilde{\mbg{\Lambda}}_{i}^{0}\mb{D}_{i}^{0} \mb{\Sigma} (\tilde{\mbg{\Lambda}}_{i}^{0}\mb{D}_{i}^{0})^\top\!\Big)^{-1}\!\!
\Delta\mbg{\lambda}_{t}\!
\right)
}
\!\right).
\label{eq:llr}
\end{align}
This statistic has the following intuition. If the alternative hypothesis is false (no outage), the expected value of \eqref{eq:llr} is in the negative domain. If the alternative hypothesis is true, the expected value of \eqref{eq:llr} is positive instead. This intuition is captured in the design of the following \cusum statistics:
\begin{align}\label{recursive_cusum}
w_t^a
=
\max\left\{0,\,
w_{t-1}^a
+
\log\!\left(\frac{f_j^{a}(\Delta\mbg{\lambda}_t)}{f_i^{0}(\Delta\mbg{\lambda}_t)}\right)
\right\},\quad \forall a\in\mathcal{A},
\end{align}
computed for every hypothesized outage.  The \cusum value $w_t^a$ evolves over time as follows. For every alternative hypothesis, the log-likelihood ratio is more likely to be negative when the hypothesis is false. Applying the max operator clips the statistic at zero, preventing accumulation of large negative values that would delay detection after a true outage occurs. In contrast, if the alternative hypothesis is true, the log-likelihood ratios are consistently positive, leading to sustained linear growth of $w_t^a$, providing evidence that the outage has occurred.

\subsection{Adapting \cusum for Bounded Demand Perturbations}

The log-likelihood ratio \eqref{recursive_cusum} assumes unbounded perturbation increments $\mbg{\xi}_t-\mbg{\xi}_{t-1}$. In practice, however, they are bounded by load limits and tend to temporally remain constant during peak-hour operation, thus violating the underlying Gaussian assumptions. Here, we propose an adaptive \cusum, which computes the likelihood only over the components of $\mbg{\xi}_t$ that are strictly within the interior of its feasible domain. 

Let $\mb{S}_{t}$ denote the selection matrix that retains only  those components of perturbation $\mbg{\xi}_t$ that remain strictly within the feasible domain at two subsequent times steps. The adapted perturbation increment and its covariance matrix are 
\begin{subequations}
\begin{align}
\Hat{\mbg{\xi}}_t - \Hat{\mbg{\xi}}_{t-1}
= \mb{S}_{t}\!\left(\mbg{\xi}_t - \mbg{\xi}_{t-1}\right) \Longrightarrow \Hat{\mbg{\Sigma}}
&= \mb{S}_{t} \mbg{\Sigma} \mb{S}_{t}^{\top}.
\label{eq:adaptive_covariance}
\end{align}
\end{subequations}

The adaptive covariance of LMP increments becomes
\begin{align*}
\mathrm{Cov}[\Delta\mbg{\lambda}_t]
&=
(\mb{S}_t\tilde{\mbg{\Lambda}}_{j}^{a}\mb{D}_{j}^{a}) \Hat{\mb{\Sigma}} (\mb{S}_t\tilde{\mbg{\Lambda}}_{j}^{a}\mb{D}_{j}^{a})^\top,\; \forall a\in\mathcal{A}\cup\{0\}.\label{eq:adapted_covarinace}
\end{align*}
Then, accommodating bounded perturbation increments, the log-likelihood ratio in \eqref{eq:llr} is computed using the adaptive covariance, while the \cusum recursion~\eqref{recursive_cusum} remains the same.

\subsection{Threshold For Outage Detection and Identification}

Running \cusum statistics in parallel for all candidate events, an outage is declared at time $\tau$ as soon as one of the statistics $\{w_t^a\}_{a\in\mathcal{A}}$ exceeds a prescribed threshold $\eta>0$: 
\begin{equation}
\tau=\inf\left\{t\geqslant1: \text{max}_{a\in\mathcal{A}} \; w_t^a\geqslant \eta\right\}.
\end{equation}
The identity of the detected outage is determined by the index $a$ of the statistic that exceeds the threshold first. As with any sequential detection procedure acting on stochastic data streams, the choice of the threshold $\eta$ governs a trade-off between detection delay and the rate of false alarms.

Rather than explicitly specifying the probability of false alarms over an infinite horizon, we adopt an average run length (ARL)-based approach for threshold selection. Specifically, under nominal operation $(\mathbb{H}_{0})$, the threshold is chosen to achieve a desired in-control ARL:
\begin{align}
    \text{ARL}_{0} = \mathbb{E}_{\mathbb{H}_0}\!\left[\tau \right], 
\end{align}
representing the expected number of samples until a false alarm is declared \cite{tartakovsky2014sequential}. This metric is estimated by evaluating the \cusum statistics on LMP streams generated from the nominal market-clearing optimization, and the threshold $\eta$ is adjusted so that the resulting $\text{ARL}_{0}$ matches a prescribed target, e.g., one false alarm per week of operation.

Once the threshold is fixed, the performance of the \cusum detector in post-outage operation $(\mathbb{H}_{1})$ is characterized in terms of the expected detection delay: 
\begin{align}
% \sup_{T \in \mathbb{N}} 
\mathbb{E}_{\mathbb{H}_1}\!\left[\tau - T \,\middle|\, \tau \geqslant T \right],
\end{align}
i.e., the expected number of post-outage samples until detection is declared. Consistent with classical sequential analysis, the detection delay scales inversely with the Kullback--Leibler divergence  $D(f^{a}\|f^0)$ between post-outage and nominal LMP distributions \cite{tartakovsky2014sequential}. Hence, outages inducing smaller distributional shifts are expected to have longer detection delays.

\subsection{Outage Identification Algorithm}
\begin{algorithm}[b]
\caption{\cusum-Based Outage Identification}
\label{alg:cusum}
\begin{algorithmic}[1]
\Statex \textbf{Input:} Sets $\mathcal{CR}_{0},\mathcal{CR}_{a\in\mathcal{A}}$, covariance $\mbg{\Sigma}$, threshold $\eta$
\Statex \textbf{Output:} Identification result $a^{\star}\in\mathcal{A}\cup\{0\}$
\State Initialize $w_0^{a} \gets 0$ for all $a\in\mathcal{A}$, $a^\star \gets 0$
\For{$t = 1,2,\dots$}
    \State Observe perturbation $\mbg{\xi}_t$ and LMP increment $\Delta\mbg{\lambda}_t$
    \State Find the critical region $\mathcal{R}_{i}^{0}\in\mathcal{CR}_{0}$ (nominal)
    \ForAll{$a\in\mathcal{A}$}
        \State Find the critical region $\mathcal{R}_{j}^{a}\in\mathcal{CR}_{a}$ (alternative)
        \State Update the \cusum statistic
        \[
        w_t^{a} \gets 
        \max\left\{0,\, w_{t-1}^{a} + \log\!\left(\frac{f_j^{a}(\Delta\mbg{\lambda}_t)}
        {f_i^{0}(\Delta\mbg{\lambda}_t)}\right)\right\}
        \]
    \EndFor
    \If{$\max_{a\in\mathcal{A}} w_t^{a} \geqslant\eta$}
        \State {\textbf{return} $a^{\star} \gets \arg\max_{a\in\mathcal{A}} w_t^{a}$}
    \EndIf
\EndFor
\end{algorithmic}
\end{algorithm}

Algorithm~\ref{alg:cusum} summarizes the outage identification procedure. It takes as input the critical regions from the nominal and hypothesized post-outage optimization problems, the demand perturbation increment covariance, and a detection threshold. It continuously outputs $a^\star=0$ unless the outage is detected, reporting outage index $a^\star \in\mathcal{A}$. At every time step, it finds the critical regions of the nominal and each alternative optimization, then updates the parallel \cusum statistics. When one of the hypothesized statistics exceeds the threshold, the algorithm declares an outage associated with that statistic.

\section{Numerical Experiments}\label{numerical_results}
The \cusum statistic is tested for line outage identification in the PJM network \cite{babaeinejadsarookolaee2019power} (Fig. \ref{fig:critical_regions}). We show how it distinguishes demand variation from outages during price spikes, then evaluate statistical performance across multiple scenarios.

\subsection{Settings}
The stochastic demand perturbation is applied to loads $1$ and $2$ as a Wiener process with a zero drift and a standard deviation of
$8$ MW, illustrated in Fig. \ref{fig:illustrative_example} (left), over a horizon of $1,000$ samples. Each sample corresponds to a market-clearing outcome computed by solving problem \eqref{prob:qp} every 5 minutes. We model a fault on the line connecting buses $1$ and $5$, which is isolated by the system operator between samples $499$ and $500$. The outage alters the multi-parametric geometry of the market-clearing problem, as depicted in Fig. \ref{fig:critical_regions} for the nominal (middle) and the post-outage (right) market operation. They respectively consist of $18$ and $9$ critical regions, which are notably different. When the same stochastic process (blue trajectories in Fig. \ref{fig:critical_regions}) develops over time, it activates distinct critical regions of the two problems, yielding an abrupt shift in the LMP streams after $T=500$, as shown in Fig. \ref{fig:illustrative_example} (middle). The proposed \cusum statistic is applied to distinguish whether these streams originate from critical regions of the nominal or post-outage problem. Detailed settings, data, and code are provided in the accompanying online repository \cite{miladpour_cusum_code}.
\begin{figure*}
    \centering
    \includegraphics[width=0.3\linewidth]{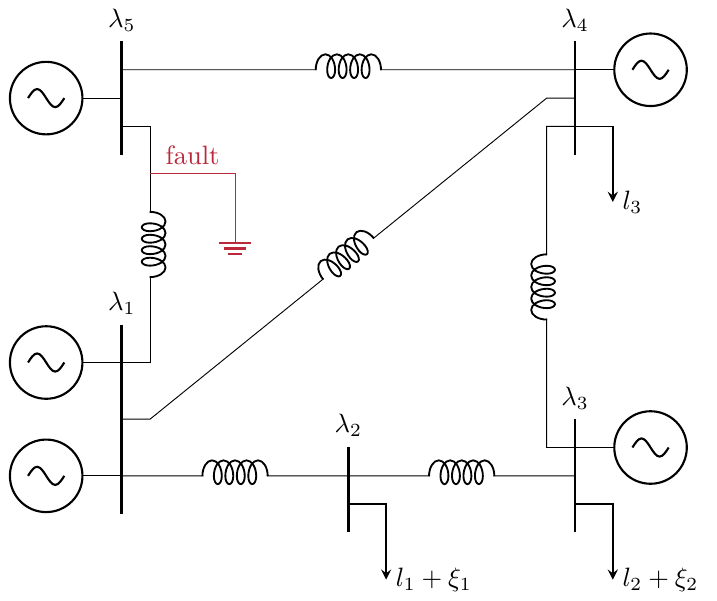}\quad\includegraphics[width=0.55\linewidth]{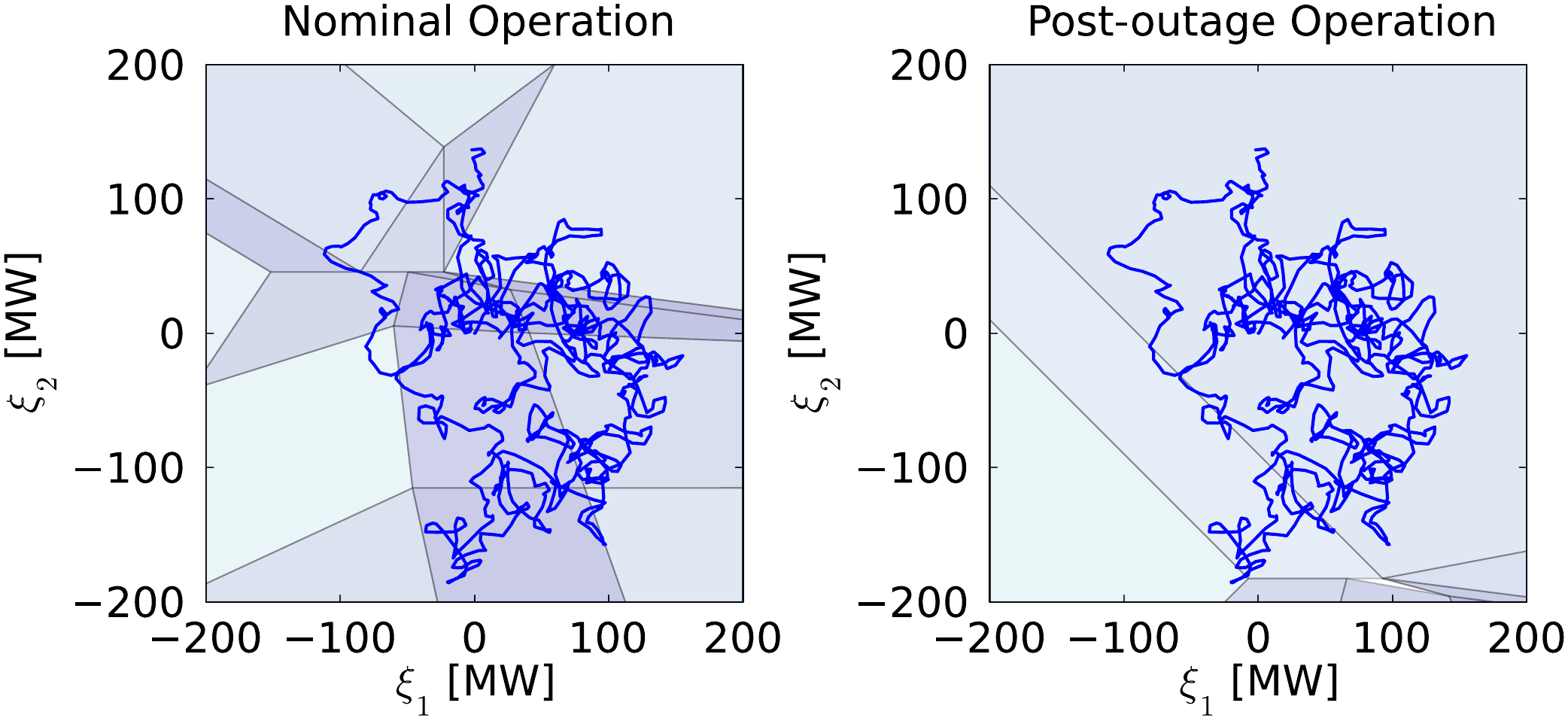}
    
    \caption{PJM system and the corresponding critical regions of the real-time market-clearing problem in the nominal and post-outage operation.}
    \label{fig:critical_regions}
\end{figure*}

\subsection{Distinguishing Line Outages from Demand Variations}

We first illustrate identification for a specific demand trajectory in Fig. \ref{fig:illustrative_example} (left), where the outage at $T=500$ coincides with a sudden demand growth. Visual inspection of demand and price trajectories might mistakenly attribute the price spike to demand growth rather than the outage. We show that the proposed identification method resolves this ambiguity.

\begin{figure*}
    \centering
    \includegraphics[width=0.95\linewidth]{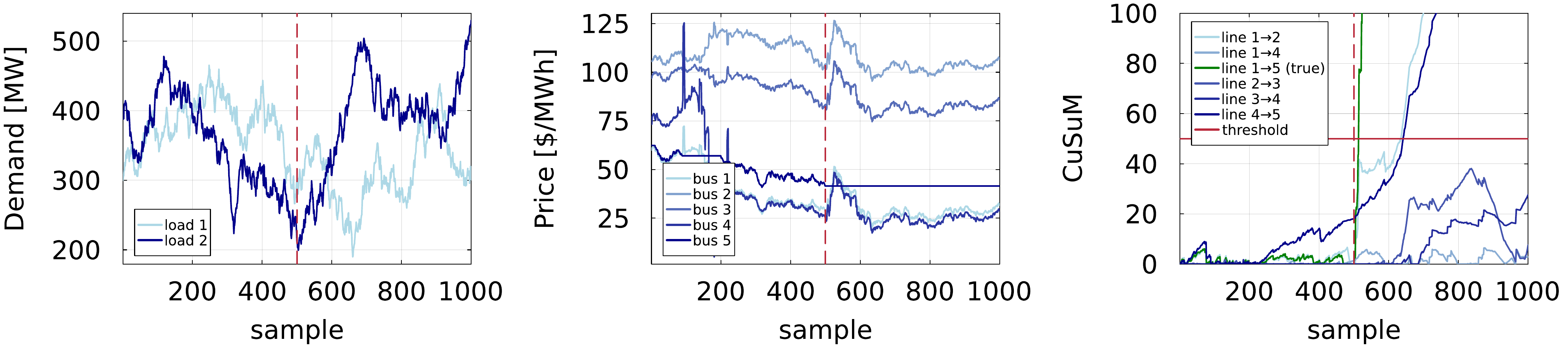}
    
    \caption{Demand (left) and LMP (middle) data streams and the dynamics of the six \cusum statistics (right). The outage occurs at $T=500$ (red dashed line).}
    \label{fig:illustrative_example}
\end{figure*}

\begin{table*}[t]
\centering
\caption{\cusum performance across 1,000 demand trajectories for varying detection thresholds $\eta$.}
\label{tab:cusum_performance}
\begin{tabular}{cccccccc}
\toprule
\multicolumn{3}{c}{Offline stage (before operation)} & \multicolumn{5}{c}{Online stage (during operation)} \\
\cmidrule(lr){1-3}\cmidrule(lr){4-8}
\multirow{3}{*}{$\eta$} & \multirow{3}{*}{\makecell{
ARL \\
$\mathbb{E}_{\mathbb{H}_0}\!\left[\tau\right]$
}} & \multirow{3}{*}{\makecell{
Probability of\\
false alarm [\%] \\
$\mathbb{P}_{\mathbb{H}_0}\!\left[a^{\star}\!\in\!\mathcal{A}\,\middle|\,\tau \!<\! t_\text{max}\right]$
}} & \multicolumn{2}{c}{\makecell{
Detection delay \\
$\tau\!-\!T \;\text{if}\; \tau\!\geqslant\!T $
}} & \multicolumn{3}{c}{Probability of $[\%]$} \\
\cmidrule(lr){4-5}\cmidrule(lr){6-8}
                 &                 &                 & average & median & \makecell{
False detection \\
$\mathbb{P}_{\mathbb{H}_1}\!\left[a^\star\!\in\!\mathcal{A}\,\middle|\,\tau \!<\! T\right]$
} & \makecell{
Successful detection \\
$\mathbb{P}_{\mathbb{H}_1}\!\left[a^\star\!\in\!\mathcal{A}\,\middle|\,\tau \!\geqslant\! T\right]$
} & \makecell{
Successful identification \\
$\mathbb{P}_{\mathbb{H}_1}\!\left[a^\star\!=\!a^{\text{true}}\,\middle|\,\tau \!\geqslant\! T\right]$
} \\
\midrule
10   & \textcolor{white}{0}871.2&90.5 & \textcolor{white}{0}56.1& 16&46.6 &53.4 & 54.5\\
20   &1793.0 & 59.1&\textcolor{white}{0}76.6 & 33&12.2 &87.7 &65.6 \\
30   &2319.2 &34.6 & 114.8& 49&\textcolor{white}{0}2.4 &97.4 &69.4 \\
40   &2732.9&15.3 & 146.1&62 &\textcolor{white}{0}0.2 &99.4 &73.1 \\
50   &3008.4 &\textcolor{white}{0}7.6 &172.7 &78 &\textcolor{white}{0}0.0 &99.4 & 76.4\\
60   &3045.8 &\textcolor{white}{0}3.0 &200.2 &93 & \textcolor{white}{0}0.0&99.4 & 78.7\\
\bottomrule
\end{tabular}
\end{table*}

We select the detection threshold $\eta=50$, whose rationale is discussed in the next subsection. Figure \ref{fig:illustrative_example} (right) depicts six \cusum statistics, each hypothesizing the outage of a specific line. Observe that, before the outage, statistics vary but hardly sustain consistent evidence accumulation. This is because the expected log-likelihood ratios are negative for all candidate hypotheses before the outage, keeping the statistics closer to zero. Line outage at $T=500$, however, triggers three out of the six statistics to suddenly grow; this is because the outage of the corresponding three lines leads to statistically similar impacts on LMPs. The statistic in green, corresponding to the true line outage, grows most rapidly and crosses the threshold first with a detection delay of only $5$ samples in this particular scenario, i.e., $25$ minutes in the real-time market timescale. This result confirms that the price spike cannot be attributed to demand variation alone, and is in fact consistent with a transmission line outage between buses $1$ and $5$. 

% \clearpage
\subsection{Performance Analysis of the \cusum Statistic}
Table~\ref{tab:cusum_performance} provides the trade-offs between performance metrics for various threshold choices $\eta$. The threshold is selected offline using a horizon of $t_\text{max}=5,000$ samples from nominal operation. Larger thresholds increase the ARL and hence reduce false alarm probability. For instance, selecting $\eta=50$ triggers a false alarm once every $\approx3,000$ samples, or once per 250 hours in real-time markets with $5$-minute resolution. 

The online performance is evaluated for the same line outage event. Larger ARLs inevitably increase detection delays, with median values ranging from $16$ samples ($1.3$h) to $93$ samples ($7.7$h). The probability of false detection is minimized for $\eta\geqslant40$, achieving near-certain detection. However, detecting a distributional shift is simpler than identifying which outage caused the shift, as shown by the probability of successful identification.  This is because multiple transmission line outages produce similar LMP distributions, yet the true outage can still be identified with probability up to $78.7$\%, but at the cost of increased detection delay. These results show that  $\eta$ can be naturally calibrated via cross-validation on known (historical) outages to balance ARL with detection delay.

\section{Conclusion}\label{conclusion}
We developed probabilistic models of real-time electricity markets and showed how they enable QCD schemes to detect and identify critical power system events from market data streams. A careful calibration of the detection threshold led to detection rates over $97\%$, while the probability of false alarm approached zero. Although identification accuracy is lower, the true outage identification rate still remains close to $79\%$. We believe that incorporating more data streams (generation by type, tie-line power flows) would improve the performance.

Our approach assumes known market parameters (e.g., generation costs), enabling precise estimation of critical regions and parametrization of probability density functions for market data streams. Future work will address parameter uncertainty and explore alternative, nonparametric QCD methods. We expect these methods may have longer detection delays, which we plan to estimate and optimize.

\newpage\clearpage
\appendix
\subsection{QP Parameters}
\label{app:qp_params}

The cost coefficients in~\eqref{prob:qp_obj} are given by
\begin{equation*}
\mb{Q} =
\begin{bmatrix}
    \mb{C} & \mb{0} \\
     \mb{0} & \mb{S}
\end{bmatrix},
\qquad
\mb{q} =
\begin{bmatrix}
    \mb{c} \\
    \mb{s}
\end{bmatrix}.
\end{equation*}
The constraint parameters in~\eqref{prob:qp_con}  are  
\setlength{\arraycolsep}{1pt}
\begin{equation*}
\mb{A} \!=\!
\begin{bmatrix*}[r]
    -\mb{1}^{\top}\mb{M}_{\text{p}} & -\mb{1}^{\top}\mb{M}_{\text{l}} \\
    \mb{F}\mb{M}_{\text{p}} & \mb{F}\mb{M}_{\text{l}} \\
    -\mb{F}\mb{M}_{\text{p}} & -\mb{F}\mb{M}_{\text{l}} \\
    \mb{I} & \mb{0} \\
    \mb{0} & \mb{I} \\
    -\mb{I} &\mb{0}  \\
     \mb{0}& -\mb{I}
\end{bmatrix*}\!\!,
\mb{B} \!=\!
\begin{bmatrix*}[r]
    -\mb{1}^{\top}\mb{M}_{\text{l}} \\
    \mb{F}\mb{M}_{\text{l}} \\
    -\mb{F}\mb{M}_{\text{l}} \\
    \mb{0} \\
    \mb{M}_{\text{l}}^{\top}\mb{M}_{\text{l}} \\
    \mb{0} \\
    \mb{0}
\end{bmatrix*}\!\!,
\mb{b} \!=\!
\begin{bmatrix*}[r]
    \mb{1}^{\top}\mb{M}_{\text{l}}\mb{l} \\
    \mb{F}\mb{M}_{\text{l}}\mb{l} + \overline{\mb{f}} \\
    -\mb{F}\mb{M}_{\text{l}}\mb{l} + \overline{\mb{f}} \\
    \overline{\mb{p}} \\
    \mb{l} \\
    -\underline{\mb{p}} \\
    \mb{0}
\end{bmatrix*}.
\end{equation*}

The auxiliary matrix to compute LMPs from the dual variables of~\eqref{prob:qp_con} is defied as 
$$\mbg{\Lambda}=\begin{bmatrix}
    \mb{1} & -\mb{F}^{\top} & \mb{F}^{\top} & \mb{0} 
\end{bmatrix},$$
yielding the standard LMP definition $$\mbg{\lambda}=\mbg{\Lambda}\mbg{\mu}=\mu_{0}\mb{1}-\mb{F}^{\top}(\overline{\mbg{\mu}}-\underline{\mbg{\mu}}),$$ where $\mu_{0}$ is the dual variable associated with the power balance constraint \eqref{eq:PDCOPF_balance}, and $\overline{\mbg{\mu}}$ and $\underline{\mbg{\mu}}$ are the dual variables of the upper and lower power flow limits in \eqref{eq:PDCOPF_flow_pos}, respectively.

\balance
\bibliographystyle{ieeetr}
\bibliography{references}

\begin{thebibliography}{10}

\bibitem{pew2015electricgrid}
{The Pew Charitable Trusts}, ``America’s electric grid: Growing cleaner, cheaper and stronger,'' October 2015.
\newblock \url{https://tinyurl.com/6ssea7ss}; Accessed: 2025-05-09.

\bibitem{ferc2021february}
{FERC}, {NERC}, and {Regional Entity}, ``{The February 2021 cold weather outages in Texas and the South Central United States},'' {\em FERC and NERC and Regional Entity Staff, Tech. Rep}, 2021.

\bibitem{li2024extreme}
F.~Li and C.~Heilbrun, ``How extreme weather events have bankrupted utility players and changed the electric grid,'' 2024.
\newblock FTI Consulting.

\bibitem{ercot2021discussion}
{ERCOT}, ``Discussion on {FERC}, {NERC}, and {NERC} regional entity staff final report,'' Dec 2021.
\newblock Available: \url{https://www.ercot.com/files/docs/2021/12/06/14_Revised_12_6_2021_ERCOT_Discussion_on_FERC,_NERC_and_NERC_Regional_Entity_Staff_Final_Report_.pdf}.

\bibitem{gridstatus}
{GridStatus.io}, ``Gridstatus: Real-time power grid status and market information.'' \url{https://www.gridstatus.io/}.

\bibitem{electricitymaps_live}
{Electricity Maps}, ``Live electricity map (15-minute intervals).'' \url{https://app.electricitymaps.com/map/live/fifteen_minutes}.

\bibitem{veeravalli2014quickest}
V.~V. Veeravalli and T.~Banerjee, ``Quickest change detection,'' in {\em Academic press library in signal processing}, vol.~3, pp.~209--255, Elsevier, 2014.

\bibitem{page1954continuous}
E.~S. Page, ``Continuous inspection schemes,'' {\em Biometrika}, vol.~41, no.~1/2, pp.~100--115, 1954.

\bibitem{chen2015quickest}
Y.~C. Chen, T.~Banerjee, A.~D. Dominguez-Garcia, and V.~V. Veeravalli, ``Quickest line outage detection and identification,'' {\em IEEE Transactions on Power Systems}, vol.~31, no.~1, pp.~749--758, 2015.

\bibitem{rovatsos2017statistical}
G.~Rovatsos, X.~Jiang, A.~D. Dom{\'\i}nguez-Garc{\'\i}a, and V.~V. Veeravalli, ``Statistical power system line outage detection under transient dynamics,'' {\em IEEE Transactions on Signal Processing}, vol.~65, no.~11, pp.~2787--2797, 2017.

\bibitem{nizam2016attack}
F.~Nizam, S.~Chaki, S.~Al~Mamun, and M.~S. Kaiser, ``Attack detection and prevention in the cyber physical system,'' in {\em 2016 international conference on computer communication and informatics (ICCCI)}, pp.~1--6, IEEE, 2016.

\bibitem{raghavan2010quickest}
V.~Raghavan and V.~V. Veeravalli, ``Quickest change detection of a markov process across a sensor array,'' {\em IEEE transactions on information theory}, vol.~56, no.~4, pp.~1961--1981, 2010.

\bibitem{tondel2003algorithm}
P.~T{\o}ndel, T.~A. Johansen, and A.~Bemporad, ``{An algorithm for multi-parametric quadratic programming and explicit MPC solutions},'' {\em Automatica}, vol.~39, no.~3, pp.~489--497, 2003.

\bibitem{taheri2020fast}
S.~Taheri, M.~Jalali, V.~Kekatos, and L.~Tong, ``Fast probabilistic hosting capacity analysis for active distribution systems,'' {\em IEEE Transactions on Smart Grid}, vol.~12, no.~3, pp.~2000--2012, 2020.

\bibitem{taheri2021strategic}
S.~Taheri, V.~Kekatos, and S.~Veeramachaneni, ``Strategic generation investment in energy markets: A multiparametric programming approach,'' {\em IEEE Transactions on Power Systems}, vol.~37, no.~4, pp.~2590--2600, 2021.

\bibitem{tate2008line}
J.~E. Tate and T.~J. Overbye, ``Line outage detection using phasor angle measurements,'' {\em IEEE Transactions on Power Systems}, vol.~23, no.~4, pp.~1644--1652, 2008.

\bibitem{tate2009double}
J.~E. Tate and T.~J. Overbye, ``Double line outage detection using phasor angle measurements,'' in {\em 2009 IEEE Power \& Energy Society General Meeting}, pp.~1--5, IEEE, 2009.

\bibitem{emami2012external}
R.~Emami and A.~Abur, ``External system line outage identification using phasor measurement units,'' {\em IEEE Transactions on Power Systems}, vol.~28, no.~2, pp.~1035--1040, 2012.

\bibitem{zhu2012sparse}
H.~Zhu and G.~B. Giannakis, ``Sparse overcomplete representations for efficient identification of power line outages,'' {\em IEEE Transactions on Power Systems}, vol.~27, no.~4, pp.~2215--2224, 2012.

\bibitem{giannakis2013monitoring}
G.~B. Giannakis, V.~Kekatos, N.~Gatsis, S.-J. Kim, H.~Zhu, and B.~F. Wollenberg, ``Monitoring and optimization for power grids: A signal processing perspective,'' {\em IEEE Signal Processing Magazine}, vol.~30, no.~5, pp.~107--128, 2013.

\bibitem{li2019real}
W.~Li, D.~Deka, M.~Chertkov, and M.~Wang, ``Real-time faulted line localization and pmu placement in power systems through convolutional neural networks,'' {\em IEEE Transactions on Power Systems}, vol.~34, no.~6, pp.~4640--4651, 2019.

\bibitem{hannon2021real}
C.~Hannon, D.~Deka, D.~Jin, M.~Vuffray, and A.~Y. Lokhov, ``{Real-time anomaly detection and classification in streaming PMU data},'' in {\em 2021 IEEE Madrid PowerTech}, pp.~1--6, IEEE, 2021.

\bibitem{kekatos2014grid}
V.~Kekatos, G.~B. Giannakis, and R.~Baldick, ``Grid topology identification using electricity prices,'' in {\em 2014 IEEE PES general meeting| conference \& exposition}, pp.~1--5, IEEE, 2014.

\bibitem{kekatos2015online}
V.~Kekatos, G.~B. Giannakis, and R.~Baldick, ``Online energy price matrix factorization for power grid topology tracking,'' {\em IEEE Transactions on Smart Grid}, vol.~7, no.~3, pp.~1239--1248, 2015.

\bibitem{birge2017inverse}
J.~R. Birge, A.~Horta{\c{c}}su, and J.~M. Pavlin, ``Inverse optimization for the recovery of market structure from market outcomes: An application to the miso electricity market,'' {\em Operations Research}, vol.~65, no.~4, pp.~837--855, 2017.

\bibitem{mitridati2017bayesian}
L.~Mitridati and P.~Pinson, ``{A Bayesian inference approach to unveil supply curves in electricity markets},'' {\em IEEE Transactions on Power Systems}, vol.~33, no.~3, pp.~2610--2620, 2017.

\bibitem{liang2023data}
Z.~Liang and Y.~Dvorkin, ``Data-driven inverse optimization for marginal offer price recovery in electricity markets,'' in {\em Proceedings of the 14th ACM International Conference on Future Energy Systems}, pp.~497--509, 2023.

\bibitem{dvorkin2020differentially}
V.~Dvorkin, P.~Van~Hentenryck, J.~Kazempour, and P.~Pinson, ``Differentially private distributed optimal power flow,'' in {\em 2020 59th IEEE Conference on Decision and Control (CDC)}, pp.~2092--2097, IEEE, 2020.

\bibitem{xie2021sequential}
L.~Xie, S.~Zou, Y.~Xie, and V.~V. Veeravalli, ``Sequential (quickest) change detection: Classical results and new directions,'' {\em IEEE Journal on Selected Areas in Information Theory}, vol.~2, no.~2, pp.~494--514, 2021.

\bibitem{roald2023power}
L.~A. Roald, D.~Pozo, A.~Papavasiliou, D.~K. Molzahn, J.~Kazempour, and A.~Conejo, ``Power systems optimization under uncertainty: A review of methods and applications,'' {\em Electric Power Systems Research}, vol.~214, p.~108725, 2023.

\bibitem{chatzivasileiadis2018lecture}
S.~Chatzivasileiadis, ``{Lecture notes on optimal power flow (OPF)},'' {\em arXiv preprint arXiv:1811.00943}, 2018.

\bibitem{casella2001statisticalinference}
G.~Casella and R.~Berger, {\em Statistical Inference}.
\newblock {Duxbury Pacific Grove}, 2002.

\bibitem{basseville1995detection}
M.~Basseville and I.~V. Nikiforov, ``Detection of abrupt changes: theory and applications,'' {\em Journal of the Royal Statistical Society-Series A Statistics in Society}, vol.~158, no.~1, p.~185, 1995.

\bibitem{tartakovsky2014sequential}
A.~Tartakovsky, I.~Nikiforov, and M.~Basseville, {\em Sequential analysis: Hypothesis testing and changepoint detection}.
\newblock CRC press, 2014.

\bibitem{babaeinejadsarookolaee2019power}
S.~Babaeinejadsarookolaee {\em et~al.}, ``{The power grid library for benchmarking AC optimal power flow algorithms},'' {\em arXiv preprint arXiv:1908.02788}, 2019.

\bibitem{miladpour_cusum_code}
M.~Hoseinpour, ``{Code for Outage Identification from Electricity Market Data: Quickest Change Detection Approach}.'' \url{https://github.com/miladpour/paper-cusum-detection}, 2026.
\newblock GitHub repository.

\end{thebibliography}
\balance
\endgroup
\end{document}